\documentclass[sigconf]{aamas}
\newcommand{\accept}{\mathrm{accept}}
\newcommand{\reject}{\mathrm{reject}}

\setcopyright{ifaamas}
\acmConference[AAMAS '21]{Proc.\@ of the 20th International Conference on Autonomous agents and Multiagent Systems (AAMAS 2021)}{May 3--7, 2021}{Online}{U.~Endriss, A.~Now\'{e}, F.~Dignum, A.~Lomuscio (eds.)}
\copyrightyear{2021}
\acmYear{2021}
\acmDOI{}
\acmPrice{}
\acmISBN{}
\usepackage{balance}
\usepackage[english]{babel}
\usepackage{tikz-cd}
\usepackage{hyperref}
\hypersetup{
    colorlinks=true,
    linkcolor=blue,
    filecolor=magenta,      
    urlcolor=cyan,
}

\usepackage{tikz}

\acmSubmissionID{???}

\title[AAMAS-2021 paper]{Nash Equilibria in Finite-Horizon Multiagent Concurrent Games}\thanks{Work supported in part by NSF grants IIS-1527668, CCF-1704883, IIS-1830549, and an award from the Maryland Procurement Office.}

\author{Senthil Rajasekaran}
\affiliation{
    \department{Computer Science Department}
    \institution{Rice University}
    }
\email{sr79@rice.edu}

\author{Moshe Y. Vardi}
\affiliation{
    \department{Computer Science Department}
    \institution{Rice University}
    }
\email{vardi@rice.edu}

\begin{abstract}
The problem of finding pure strategy Nash equilibria in multiagent concurrent games with finite-horizon temporal goals has received some recent attention. Earlier work solved this problem through the use of Rabin automata. In this work, we take advantage of the finite-horizon nature of the agents' goals and show that checking for and finding pure strategy Nash equilibria can be done using a combination of safety games and lasso testing in B\"uchi automata. To separate strategic reasoning from temporal reasoning, we model agents' goals by deterministic finite-word automata (DFAs), since finite-horizon logics such as LTL\textsubscript{f} and LDL\textsubscript{f} are reasoned about through conversion to equivalent DFAs. This allows us to characterize the complexity of the problem as PSPACE complete.
\end{abstract}
\begin{document}
\pagestyle{fancy}
\fancyhead{}
\settopmatter{printfolios=true}
\maketitle 
\section{Introduction}
Game theory provides a powerful framework for modeling problems in system design and verification \cite{Henzinger05, LogicGames, GTW02}. In particular, two-player games have been used in synthesis problems for temporal logics \cite{PnuRos89a}. In these games, one player takes on the role of the system that tries to realize a property and the other takes on the role of the environment that tries to falsify the property. Within the scope of multiplayer games, two-player zero-sum games are the easiest to analyze, since they are purely adversarial -- there is no reason for either player to do anything but maximize their own utility at the expense of the other.

When there are multiple agents with multiple goals, pure antagonism is not a reasonable assumption \cite{Wool11}. \emph{Concurrent games} are a fundamental model of such multiagent systems \cite{alur2002alternating,mogavero2014reasoning}. \emph{Iterated Boolean Games} (iBG) \cite{iBG} are a restriction of concurrent games introduced in part to generalize temporal synthesis problems to the multiagent setting. In an iBG, each agent has a temporal goal, usually expressed in \emph{Linear Time Temporal Logic} (LTL) \cite{Pnu77}, and is given control over a unique set of boolean variables. At each time step, the agents collectively decide a setting to all boolean variables by individually and concurrently assigning values to their own variables. This creates an infinite sequence of boolean assignments (a \emph{trace}) that is used to determined which goals are satisfied and which are not \cite{iBG}. In this paper, we generalize the iBG formalism slightly to admit arbitrary finite alphabets rather than just truth assignments to boolean variables, as discussed below.

The concept of the \emph{Nash Equilibrium} \cite{Nash48} is widely accepted as an important notion of a solution in multiagent games and represents a situation where agents cannot improve their outcomes unilaterally. In this paper we consider deterministic agents, and therefore the notion of a Nash equilibrium in this paper that of pure
strategy Nash equilibrium \cite{SLmultiagentbook}. 
This definition has a natural analogue when iBGs are considered, so finding Nash Equilibria in iBGs is an effective way to reason about temporal interactions between multiple agents \cite{iBG}.  This problem has received attention in the literature when the goals are derived from infinite-horizon logics such as LTL \cite{FKL10,GutierrezNPW20}. There are, however, interactions that are better modeled by finite-horizon goals, especially when notions such as ``completion'' are considered \cite{GV15}. In such settings, it is more effective to reason about goals that can be completed in some finite but perhaps unbounded number of steps. Thus, while the agents still create an infinite trace with their decisions, satisfaction occurs at a finite time index.  With this modification in mind, the analogous problem for finite-horizon temporal logics has recently began to receive attention \cite{GPW17}. The main result of \cite{GPW17} is that automated equilibrium analysis of finite-horizon goals in iterated Boolean games can be done via reasoning about automata on infinite words, specifically, \emph{Rabin automata}. 

Here we address a more abstract version of the multi-agent finite-horizon temporal-equilibrium problem by analyzing concurrent iterated games in which each agent is given their own \emph{Deterministic Finite Word Automata} (DFA) goal. The reason for this is twofold. First, essentially all finite-horizon temporal logics are reasoned about through conversion to equivalent DFA, including the popular logics LTL\textsubscript{f} and  LDL\textsubscript{f} \cite{GV13,GV15}. Thus, using DFA goals offers us a general way of dealing with a variety of temporal formalisms. Furthermore, using DFA goals enables us to separate the complexity of temporal reasoning from the complexity of strategic reasoning. Our focus on DFAs also ties in to a growing interest in DFAs as graphical models that can be reasoned about directly in a number of related fields; see \cite{MSVBCP19, HJATD, Y18} for a few examples in the context of machine learning. 

Our modelling of this problem is done from the viewpoint of a system planner. Specifically, when given a system in which multiple agents have DFA goals, we query a subset $W$  of ``good'' agents to see if there is Nash equilibrium in which only the agents in $W$ are able to satisfy their goals. By the definition of the Nash equilibrium, this means that agents not within $W$, which  we consider as ``bad'' agents, are unable to unilaterally change their strategy and satisfy their own ``bad'' goal. In doing so we can naturally incorporate malicious agents with goals contrary to the planner's by specifying a set $W$ that not contain such agents. This study of teams of cooperating agents has clear parallels to earlier work in rational synthesis \cite{FKL10,KupfermanPV16}.

Our main result is that automated temporal-equilibrium analysis is PSPACE complete. We prove that the problem of identifying sets of players that admit Nash equilibria in concurrent multi-agent games with DFA goals can be solved using rather simple constructions. Specifically, our algorithm works by first solving a safety game for each agent in the game and then considers nonemptiness in a B\"uchi word automata constructed with respect to the set $W$ of agents, which can be done in PSPACE. %We analyze the computational complexity of the problem, and show that checking whether a set $W$ of agents has a Nash Equilibrium is PSPACE-complete. 
This is in contrast to the 2EXPTIME upper bound of \cite{GPW17}, which analyzed the combined complexity of temporal \emph{and} strategic reasoning and also considered existence overall instead of with respect to a specific set of agents $W$. In this case driving force behind the complexity result was the doubly exponential blow-up from LDL\textsubscript{f} to DFAs \cite{GV13,KV01d}. Finally, we prove our algorithm optimal by providing a matching lower bound.

% Using DFA goals allows us to pinpoint the complexity of analyzing strategic behavior,  as translating from LTL\textsubscript{f} or  LDL\textsubscript{f} to DFAs is known to be a doubly exponential blow-up \cite{GV13,KV01d}. Furthermore, the analysis technique in terms of safety and reachability is significantly simpler than the analysis in terms Rabin automata in \cite{GPW17} and has the potential for a more scalable implementation. Lastly, we are able to prove this algorithm optimal by providing a tight lower bound.
%MYV1:
%including implementations in the form of symbolic algorithms. Third, we prove our algorithm optimal by providing a tight lower bound.

\section{Background}

\subsection{Automata Theory}

We assume familiarity with basic automata theory, as in \cite{sipser2006}. Below is a quick refresher on $\omega$-automata and infinite tree automata.

%MYV1: Please see how I compressed this.
\begin{definition} [$\omega$ automata]\cite{GTW02}
A deterministic  $\omega$ automaton is a 5-tuple $\langle Q, q_0 , \Sigma, \delta,  Acc \rangle $, where $Q$ is a finite set of states,  $q_0 \in Q$ is the initial state, $\Sigma$ is a finite alphabet, $\delta: Q \times \Sigma \rightarrow Q$ is the transition function, and 
$Acc$ is an acceptance criterion. An infinite word $w=a_0,a_1,\ldots\in \Sigma^\omega$ is accepted by the automaton, if the run $q_0,q_1, \ldots \in Q^{\omega}$ is accepting, which requires that $q_0$ is the initial state and $q_{i+1} = \delta(q_i, a_i) $ for all $i\geq 0$, The run $q_0,q_1 \ldots$ satisfies the acceptance condition $Acc$.
\end{definition}
 
\begin{definition}[$\omega$ automata B\"uchi Acceptance Condition]\cite{GTW02}
The B\"uchi condition is specified by a finite set $F \subseteq Q$.
For a given infinite run $r$, let $inf(r)$ denote the set of states that occur infinitely often in $r$.  We have that the B\"uchi condition is satisfied by $r$ if $inf(r) \cap F \not = \emptyset$
\end{definition} 

We now extend this definition to deterministic B\"uchi tree automata. These automata will recognize a set of \emph{labeled} directed trees.
A $\Sigma$-labeled, $\Delta$-directed tree, for finite alphabets $\Sigma$ (\emph{label alphabet}, or \emph{labels}, for short) and $\Delta$ (\emph{direction alphabet}, or \emph{directions} for short) is a mapping $\tau:\Delta^*\rightarrow \Sigma$. Intuitively, $\tau$ labels the nodes $u\in\Delta^*$ with labels from $\Sigma$. A \emph{path} $p$ of a $\Delta$-directed tree is an infinite  sequence $p=u_0,u_1,\ldots\in (\Delta^*)^\omega$, such that $u_{i+1}=u_i b_i$ for some $b_i\in\Delta$. We use the notation $\tau(p)$ to denote the infinite sequence $\tau(u_0),\tau(u_1),\ldots \in \Sigma^\omega$. 
\begin{definition}[Deterministic B\"uchi Tree Automata]\cite{GTW02}
A deterministic B\"uchi tree automaton is a tuple $\langle \Sigma, \Theta, Q,  q_0, \Delta, F \rangle$, where $\Sigma$ is a finite label alphabet, $\Delta$ is a finite direction alphabet, $Q$ is a finite state set, $q_0\in Q$ is the initial state, $\rho: (Q \times \Sigma \times \Delta) \rightarrow Q$ is a deterministic transition function, and  $F\subset Q$ is the accepting-state set.

The automaton is considered to be top-down if runs of the automata start from the root of a tree. All automata in this paper will be top-down, and our notion of a run is conditioned on this.

A run of this automaton on a $\Sigma$-labeled, $\Delta$-directed tree $\tau: \Delta^* \rightarrow \Sigma$ is a $Q$-labeled, $\Delta$-directed tree $r: \Delta^* \rightarrow Q$ such that $r(\varepsilon)=q_0$, and if $u\in\Delta^*$, $\tau(u)=a$, for $a\in\Sigma$, $r(u)=q$, and $v=ub$ for $b\in\Delta$, then $r(v)=\rho(q,a,b)$. The run $r$ is accepting if $r(p)$ satisfies the B\"uchi condition $F$ for every path $p$ of $r$.

\end{definition} 

\subsection{Games}

% \textbf{This section should be cut a lot as well}
In this section we provide some definitions related to simple two player games to provide a standard notation throughout this paper. The two players will be denoted by player $0$ and player $1$.
\begin{definition} [Arena]

An \emph{arena} is a four tuple $A = (V, V_0, V_1, E)$ where $V$ is a finite set of vertices, $V_0$ and $V_1$ are disjoint subsets of $V$ with $V_0 \cup V_1 = V$ that represent the vertices that belong to player $0$ and player $1$ respectively, and $E \subseteq V \times V$ is a set of directed edges, i.e. $(v, v') \text{ } | \text{ } \in E$ if there is an edge from $v$ to $v'$.

Intuitively, the player that owns a node decides which outgoing edge to follow.
Since $V = V_0 \cup V_1$, we can notate the same arena while omitting $V$, a convention we follow in this paper.
\end{definition}

\begin{definition}[Play]
A \emph{play} in an arena $A$ is an infinite sequence $\rho_0 \rho_1 \rho_2 \ldots \in V^{\omega}$ such that $(\rho_n , \rho_{n+1}) \in E$ holds for all $n \in \mathbb{N}$. We say that $\rho$ starts at $\rho_0$
\end{definition}

We now introduce a very broad definition for two-player games.
\begin{definition}[Game]
A \emph{game} $G = (A,Win)$ consists of an arena $A$ with vertex set $V$ and a set of winning plays $Win \subseteq V^{\omega}$. A play $\rho$ is winning for player $0$ if $\rho \in Win$, otherwise it is winning for player $1$.

\end{definition}

Note that in this formulation of a game, reaching a state $v \in V$ with no outgoing transitions is always losing for player $0$, as player $0$ is the one that must ensure that $\rho$ is infinite ( a member of $V^{\omega}$).

 A game is thus defined by its set of winning plays, often called the winning condition. One such widely used winning condition the safety condition.

\begin{definition}[Safety Condition/ Safety Game]

Let $A = (V, V_0, V_1,\\ E)$ be an arena and $S \subseteq V$ be a subset of $A$'s vertices. Then, the \emph{safety condition} $Safety(S)$ is defined as  $Safety(S) = \{ \rho \in V^{\omega} \text{ } | \text{ } Occ(\rho) \subseteq S \}$
where $Occ(\rho)$ denotes the subset of vertices that occur at least once in $\rho$. 

A game with the safety winning condition for a subset $S$ is a \emph{safety game} with the set $S$ of safe vertices. Information about solving safety games, including notions of \emph{winning strategies} and \emph{winning sets} can be found here \cite{McNaughton1993InfiniteGP}.

\end{definition}

\subsection{Concurrent Games and iBGs}
A \emph{concurrent game structure} (CGS) is an 8-tuple 
$$( Prop, \Omega, (A_i)_{i \in \Omega}, S, \lambda, \tau, s_0 \in S, (A^i)_{i \in \Omega})$$ %inlining is just too ugly, I'll come back here as a last resort
where $Prop$ is a finite set of \emph{propositions}, $\Omega =\{ 0, \ldots k-1 \} $ is a finite set of \emph{agents}, $A_i$ is a set of \emph{actions}, where each $A_i$ is associated with an agent $i$ (we also construct the set of \emph{decisions}  $D =  A_0 \times A_1 \ldots A_{k- 1}$,  $S$ is a set of \emph{states}, $\lambda: S \rightarrow 2^{Prop} $ is a \emph{labeling function} that associates each state with a set of propositions that are interpreted as true in that state, $\tau : S \times D \rightarrow S$ is a deterministic \emph{transition function} that takes a state and a decision as input and returns another state, $s_0$ is a state in $S$ that serves as the \emph{initial state}, and $A^i$ is a DFA associated with agent $i$. A DFA $A^i$ is denoted as the goal of agent $i$. Intuitively, agent $i$ prefers plays in the game that satisfy $A^i$, that is a play such that some finite prefix of the play is accepted by $A^i$. It is for this reason we refer to $A^i$ as a "goal".

We now define \emph{iterated boolean games} (iBG), a restriction on the CGS formalism. Our formulation is slight generalization of the iBG framework introduced in \cite{iBG}, as we take the set of actions to be a finite alphabet rather than a set of truth assignments since we are interested in separating temporal reasoning from strategic reasoning. An iBG is defined by applying the following restrictions to the CGS formalism. Each agent $i$ is associated with its own alphabet $\Sigma_i$. These $\Sigma_i$ are disjoint and each $\Sigma_i$ serves as the set of actions for agent $i$; an action for agent $i$ consists of choosing a letter in $\Sigma_i$. The set of decisions is then $\Sigma = \bigtimes_{i=0}^{k-1} \Sigma_i$. The set of states corresponds to the set of decisions $\Sigma$; there is a bijection between the set of states and the set of decisions. The labeling function mirrors the element of $\Sigma$ associated with each state. As in \cite{iBG}, we still have $\lambda(s) = s$, but with $s \in \Sigma$ now. As a slight abuse of notation, we consider the ``proposition'' $\sigma \in \Sigma_i$ for some $i$ to be true at state $s$ if $\sigma$ appears in $s$, allowing us to generalize towards arbitrary alphabets. Finally, the transition function $\tau$ is simply right projection $\tau(s,d) = d$.

We now introduce the notion of a \emph{strategy} for agent $i$ in the general CGS formalism.
\begin{definition}[Strategy for agent $i$]
A strategy for agent $i$ is a function $\pi_i : S^* \rightarrow A_i$. Intuitively, this is a function that, given the observed history of the game (represented by an element of $S^*$), returns an action $a_i \in A_i$.
\end{definition}

Recalling that $\Omega = \{ 0, 1 \ldots k-1 \}$ represents the set of agents, we now introduce the notion of a \emph{strategy profile}.
\begin{definition}[Strategy Profile]
Let $\Pi_i$ represent the set of strategies for agent $i$. Then, we define the set of strategy profiles  
$ \Pi = \bigtimes_{i \in \Omega} \Pi_i$
\end{definition}
Note that since both the notion of strategies for individual agents and the transition function in a CGS are deterministic, a given strategy profile for an CGS defines a unique element of $S^{\omega}$ (a trace).

\begin{definition}[Primary Trace resulting from a Strategy Profile]
Given a strategy profile $\pi$, the primary trace of $\pi$ is the unique trace $t$ that satisfies
\begin{enumerate}
    \item $t[0] = \pi(\epsilon)$
    \item $t[i] = \pi( t[0], \ldots t[i-1])$
\end{enumerate}
We denote this trace as $t_{\pi}$.
\end{definition}

Given a trace $t \in S^\omega$, define the \emph{winning set} $W_t=\{i\in\Omega~:~ t\models A^i \}$ to be the set of agents whose DFA goals are satisfied by a finite prefix of the trace $t$. The \emph{losing set} is then defined as $ \Omega / W_t$.

A common solution concept in game theory is the \emph{Nash equilibrium}, which we will now modify to fit our iBG framework. In our framework, a Nash equilibrium is a strategy profile $\pi$ such that for each agent $i$, if $A^i$ is not satisfied on $t_{\pi}$, then any unilateral strategy deviation for agent $i$ will not result in a trace that satisfies $A^i$. Formally:
 \begin{definition}[Nash Equilibrium]\cite{iBG}
Let $G$ be an iBG and $\pi = \langle \pi_0 , \pi_1 \ldots  \pi_{k-1}\rangle$ be a strategy profile. We denote $W_{\pi}=W_{t_\pi}$.  The profile $\pi$ is a \emph{Nash equilibrium} if for every $i \in \Omega / W_t$ we have that given all strategy profiles of the form $\pi' = \langle\pi_0 , \pi_1 \ldots \pi'_i \ldots  \pi_{k-1}\rangle$, for every $\pi^{'}_i \in \Pi_i$, it is the case that $i \in \Omega / W_{\pi '}$.
\end{definition} 

This definition provides an analogy for the Nash equilibrium defined in \cite{Nash48} by capturing the same property - no agent can unilaterally deviate to improve its own payoff (moving from having a not satisfied goal to a satisfied goal). Agents already in the set $W_{\pi}$ cannot have their payoff improved further, so we do not check their deviations.

Our paper is based around one central question: \emph{Given an iBG, which subsets of agents admit at least one Nash equilibrium?}

\section{Tree Automata Framework}

In order to address our central question, we first describe a tree-automata framework to characterize the set of Nash equilibrium strategies in an iBG $G$. In this section we fix a winning set $W \subset \Omega$ and then describe a deterministic B\"uchi tree automaton that recognizes the set of strategy profiles for $W$. In the next section we develop an algorithm based on this tree-automata framework.

Given $k$ DFA goals corresponding to $k$ agents, we retain the notation that the set of actions for agent $i$ is given by $\Sigma_i$. The goal DFA for agent $i$ will then denoted as $A^i = \langle Q^i, q_0^i, \Sigma, \delta^i, F^i\rangle$. Note that the alphabet of the DFA is $\Sigma$, since it transitions according to decisions by \emph{all} agents in the overlying iBG structure. Since $\Sigma=\Sigma_0 \times \ldots \Sigma_{k-1}$, compact notation is often used to describe the transition function $\delta^i$. For example, the {\sc Mona} tool uses binary decision diagrams to represent automata with large alphabets \cite{EKM98}. 

\subsection{Strategy Trees and Tree Automata}
As defined previously, strategy profiles are functions $ \pi :\Sigma^* \rightarrow \Sigma$. Therefore, strategy profiles correspond exactly towards labeled $\Sigma$-labeled trees, which are defined in the exact same way. We use the common notions of tree paths and label-direction pairs as widely defined in the literature (see \cite{GTW02} for reference).

A $W$-NE-strategy, for $W\subseteq\Omega$,  is a mapping
$\pi: \Sigma^* \rightarrow \Sigma$ such that the following conditions are satisfied:
\begin{enumerate}
\item {\sf Primary-Trace Condition}: The primary infinite trace $t_\pi$ defined by $\pi$ satisfies the goals $A^j$ precisely for $j\in W$. The trace $t_\pi=x_0,x_1,\ldots$ for $\pi$ is once again defined as follows
\begin{enumerate}
\item 
$x_0=\varepsilon$
\item
$x_{i+1}=x_0,\ldots,x_i,\pi(x_0,\ldots,x_i)$
\end{enumerate}
\item
{\sf $j$-Deviant-Trace Condition}:
Each $j$-\emph{deviant} trace $t=y_0,y_1,\\ \ldots$ for $j\not\in W$, does not satisfy the goal $A^j$.

For $\alpha \in \Sigma$,  we introduce the notation 
$\alpha[-j]$ to refer to $\alpha|_{\Sigma \setminus \Sigma_j}$ (that is, $\alpha$ with $\Sigma_j$ projected out). A trace $t=y_0,y_1,\ldots$ is $j$-deviant if
\begin{enumerate}
\item
$y_0=\varepsilon$
\item
$y_{i+1}=y_0,\ldots,y_i,\alpha$, where $\alpha\in \Sigma$ and 
$\alpha[-j]=\pi(y_i)[-j]$
\item $t$ is not the primary trace
\end{enumerate}
\end{enumerate}

In order to simplify the presentation, we introduce the assumption that for all agents $j$ we have $|\Sigma_j| \geq 2$. This is because there are no $j$-Deviant-Traces for an agent with only one strategy. Therefore $W$-NE analysis only amounts to checking the Primary-Trace Condition for these agents.

Note that there are traces that do not fall into either category. For example, we could have a trace that contains a label direction pair $(\alpha,\beta)$ such that $\alpha[-j] \not = \beta[-j]$ for all $j \in \Omega \setminus W$. Or, we could have a trace that contains two label direction pairs $(\alpha_1,\beta_1)$ and $(\alpha_2,\beta_2)$ such that $\alpha_1 \not = \beta_1$, $\alpha_1[-j_1] = \beta[-j_1]$, $\alpha_2 \not = \beta_2$, and $\alpha_2[-j_2] = \beta[-j_2]$ for $j_1\not=j_2$. Traces like these and others that do not fit into either the Primary-Trace category or the $j$-Deviant-Trace category are irrelevant to the Nash equilibrium condition -  it does not matter what properties do or do not hold on these traces. As a reminder, a trace $z_0,z_1,\ldots \in \Sigma^{\omega}$ satisfies a DFA $A$ if $A$ accepts $z_0,\ldots,z_k$ for some $k\geq 0$.

To check if there exists a $W$-NE strategy, we construct an infinite-tree automaton $T$ that accepts all $W$-NE strategies. The problem of determining whether a $W$-NE exists then reduces to querying $L(T)\not=\emptyset$.  Recall that we notate the goal DFA of agent $i$ as $A^i = \langle Q^i, q_0^i, \Sigma, \delta^i, F^i \rangle$. We assume that that $q^i_0\not\in F^j$, since we are not interested in empty traces. We first construct a deterministic B\"uchi automaton $A_W=\langle Q,q_0,\Sigma,\delta,F \rangle$ that accepts a word in $\Sigma^\omega$ if it satisfies precisely the goals $A^j$ for $j\in W$. Intuitively, $A_W$ simulates concurrently all the goal DFAs, and checks that $A^j$ is satisfied precisely for $j\in W$. We define the following for $A_W$.
\begin{enumerate}
\item
$Q=(\bigtimes_{j\in\Omega} Q^j) \times 2^\Omega$ 
\item
$q_0=\langle q^1_0,\ldots,q^n_0,W \rangle$
\item
$F=(\bigtimes_{j\in\Omega} Q^j) \times \{\emptyset\}$
\item
$\delta(\langle q_1,\ldots q_n,U\rangle,\alpha)=\langle q'_1,\ldots q'_n,V\rangle$ if
\begin{enumerate}
\item $q'_j=\delta^i(q_j,\alpha)$, 
where $q'_j\not\in F^j$ for $j\not\in W$, and 
 $V=U-\{j: q_j'\in F^j\}$.
\end{enumerate}
\end{enumerate}
Note that $A_W$ concurrently simulates all the goal DFAs while it also checks that no goal DFA $A^j$ for $j\not\in W$ is satisfied. (Note that if $q'_j\in F^j$ for $j\not\in W$, then the transition is not defined, and $A_W$ is stuck.) The last component of the state holds the indices of the goals that are yet to be satisfied. For $A_W$ to accept an infinite trace, all goals $A^j$ for $j\in W$ have to be satisfied, so the last component of the state has to become empty. Note that if $A_W$ reaches an accepting state in $F$, then it stays in the set $F$ unless it gets stuck. 

\begin{lemma}[$A_W$ Correctness]
For a given $W \subseteq \Omega$, the automaton $A_W$ accepts an $\omega$-word $u \in {\Sigma}^{\omega}$ iff $u\models A^i$ for precisely the agents $i \in W$.
\end{lemma}

\begin{proof}
First, note that no prefix of $w$ can satisfy $A^j$ for some $j \in \Omega \setminus W$. If that is the case, then by the definition of the transition function $\delta$ we would have no transition defined upon reading this prefix, meaning that $A_W$ cannot accept. Next, note that every goal $A^j$ for $j \in W$ must be satisfied by a prefix of $w$. Otherwise, the $2^{\Omega}$ component of the states in $Q$ would never reach $\emptyset$, as the only way to remove elements from this component is to satisfy the goals $A^j$ for $j \in W$. Since the B\"uchi acceptance condition implies that a final state in $A_W$ be reached, we know that when a final state is reached all goals $A^j$ for $j \in W$ have previously been satisfied. Since both of these conditions must hold, we conclude the lemma.
\end{proof}

We now construct a deterministic top-down B\"uchi tree automaton $T_0$ that accepts an infinite tree $\pi: \Sigma^*\rightarrow \Sigma$ if the Primary-Trace Condition with respect to $W$ holds. Essentially, $T_0$ runs $A_W$ on the primary trace defined by the input strategy $\pi$. Formally, $T_0=(\Sigma,\Sigma,Q\cup\{q_a\},q_0,\rho_0,F\cup\{q_a\})$, where:
\begin{enumerate}
\item 
$\Sigma$ is both the label alphabet of the tree and its set of directions, Here we introduce the notation that $\alpha$ is an element of $\Sigma$ corresponding to a label and $\beta$ is an element of $\Sigma$ corresponding to a direction.
\item
$q_a$ is a new accepting state
\item
For a state $q$, label $\alpha$, and direction $\beta$, we have $\rho_0(q,\alpha,\beta)=\delta(q,\alpha)$ if $\alpha=\beta$ and $q \not = q_a$, and  $\rho_0(q,\alpha,\beta)=q_a$ otherwise 
\end{enumerate}
Note that $T_0$ simulates $A_W$ along the branch corresponding to the primary trace defined by the input tree $\pi$. Along all other branches, $T_0$ enters the accepting state $q_a$.  

\begin{lemma}
Let $G$ be an iBG and $W \subseteq \Omega$ be a set of agents. Let $\pi: \Sigma^*\rightarrow \Sigma$ be a strategy profile. Then $\pi$ is accepted by the tree automaton $T_0$ iff $\pi$ satisfies the Primary Trace condition.
\end{lemma}

\begin{proof}
The primary trace is a single path $p \in \pi$ such that for all label direction pairs $(\alpha,\beta) \in p$ we have $\alpha = \beta$. The automata $T_0$ transitions to the state $q_A$ immediately after seeing a label-direction pair $(\alpha,\beta)$ such that $\alpha \not = \beta$, meaning that acceptance by $T_0$ is solely determined by acceptance on the path $p$ with $\alpha = \beta$ for every $(\alpha,\beta) \in p$, which is the primary trace of $\pi$ by definition.

The Primary-Trace Condition is that on the primary trace, only the goals $A^i$ for $i \in W$ are satisfied. By virtue of construction, $T_0$ simulates the DBW $A_W$ on the primary trace, which captures this condition by the previous arguments presented in the construction of $A_W$ in Lemma 3.1.
\end{proof}

We also construct a deterministic top-down B\"uchi infinite-tree automaton $T_j$ that accepts precisely the trees $\pi : \Sigma^* \rightarrow \Sigma$ that satisfy the $j$-Deviant-Trace Condition. Given a DFA goal $A^j=(Q^j,q^j_0,\Sigma,\delta^j,F^j)$, we define 
$T_j=(\Sigma,\Sigma, (Q^j \times \{0,1\}) \cup \{q_A \} ,\langle q^j_0,0\rangle, \\ \rho_j,(Q^j \times \{0\}) \cup ((Q^j \setminus F^j)\times \{1\}) \cup \{q_A \})$, where:
\begin{enumerate}
    \item $\Sigma$ is both the label alphabet of the tree and its set of directions. We retain the notation that $\alpha$ is a label and $\beta$ is a direction.
    \item $q_A$ is a new accepting state. (By a slight abuse of notation we consider $q_A$ to be a pair $\langle q_A,0\rangle$.) 
    \item We maintain two copies of $Q^j$, one tagged with 0 and one tagged with 1. Intuitively, we stay in $Q^j\times\{0\}$ on the primary trace until there is a $j$-deviation, and then we transition to $Q^j\times\{1\}$,
    \item
        
    $\rho_j( \langle q,i\rangle, \alpha, \beta)$ is defined as follows
    \begin{enumerate}
    \item $\delta^j(q,\alpha) \times \{0\}$  \text{if $i=0$ and $\alpha = \beta$ } 
    
    \item $\delta^j(q, \beta) \times \{1\}$  \text{if $i=0$, $\alpha\not=\beta$, $\alpha[-j] = \beta[-j]$ 
    
    and $ \delta^j(q,\beta) \not \in F^j$}
    
    \item $\delta^j(q, \beta) \times \{1\}$ \text{if $i=1$, $\alpha[-j] = \beta[-j]$, and $\delta^j(q,\beta) \not \in F^j$}
    
    \item $q_A$ \text{if $q = q_A$ or $\alpha[-j] \not = \beta[-j]$ } 
    \end{enumerate}

\end{enumerate}

On the primary trace of $\pi$, we enter states $q \in Q^j \times \{0\}$. All of these states are accepting, so the primary trace will always be an accepting branch in $T_j$ since the primary trace is not relevant to the $j$-Deviant-Trace Condition. 
Intuitively, we may leave the primary trace at a node labeled $\alpha$ by following a direction $\beta$ such that $\alpha[-j] = \beta[-j]$ and $\alpha \not = \beta$. Here, we transition to the second copy of $Q^j$, $Q^j \times \{1\}$, where the $1$ denotes that we have left the primary trace. When we are in these states on a node labeled $\alpha$, we may transition according to $\delta^j$ on any direction $\beta$ with $\beta[-j] = \alpha[-j]$. Nevertheless, due to how the transitions are defined, we can never enter a state in $F^j$. If such a direction $\beta$ exists such that $\beta[-j] = \alpha[-j]$ and the resulting transition according to $\delta^j$ would put $A^j$ in $F^j$, then the automaton does not have a defined transition and therefore can not accept on this path. Otherwise, if we see a direction $\beta$ such that for our current label $\alpha$ we have that $\alpha[-j] \not = \beta[-j]$, then this no longer corresponds to a $j$-Deviant-Trace. At this point we transition to $q_A$, a catch-all accepting state that marks all continuations of the current path irrelevant to the $j$-Deviant-Trace Condition. Therefore if we are in state $q_A$ we transition back to $q_A$ on all directions $\beta$ regardless of the label.

\begin{lemma}
Let $G$ be an iBG and $W \subseteq \Omega$ be a set of agents. Let $\pi: \Sigma^*\rightarrow \Sigma$ be a strategy profile. Then $\pi$ is accepted by the tree automaton $T_j$ iff $\pi$ satisfies the $j$-Deviant Trace Condition.
\end{lemma}

\begin{proof}
By definition set of $j$-Deviant-Traces is the set of paths $p$ such that for all $(\alpha,\beta) \in p$ we have $\alpha[-j] = \beta[-j]$, excluding the primary trace. The $j$-Deviant-Trace Condition says that for all these paths $p$, none have a finite prefix accepted by $A^j$.

For a infinite/finite sequence of label direction pairs $p = (\alpha_0, \beta_0) \\ \ldots$, let $\beta_p$ denote the infinite/finite word obtained by concatenating all $\beta$ together in index order. If $\pi$ does not satisfy the $j$-Deviant-Trace Condition, then there exists a finite sequence of label-direction pairs $p_j =  (\alpha_0, \beta_0) \ldots (\alpha_n, \beta_n)$ such that $\forall i ( 0 \leq i  \leq n) . \alpha_i[-j] = \beta_i[-j]$, $\exists i  (0 \leq i  \leq n) . \alpha_i \not = \beta_i$,  and $A^j$ accepts $\beta_{p_j}$.  Since $\alpha_i[-j] = \beta_i[-j]$ for every index in $p_j$, $T_j$ never attempts to transition to $q_A$ along $p_j$. And since $A^j$ accepts $\beta_{p_j}$, we know that along $p_j$ $T_j$ attempts to transition to a final state in $F^j$ and get stuck, therefore rejecting. Therefore, $T_j$ reject $\sigma$.

Now assume that $T_j$ does not accept $\pi$. This means that along some $j$-Deviant-Trace $T_j$ attempts to transition to a state $q_f \times \{ 1 \}$, where $q_f \in F^j$, and gets stuck, as this is the only way for $T_j$ to reject. This follows from the observation that every reachable state in $T_j$ is accepting. This means there exists a finite sequence of label-direction pairs $p_j =  (\alpha_0, \beta_0) \ldots (\alpha_n, \beta_n)$ such that $\forall ( 0 \leq i  \leq n ). \alpha_i[-j] = \beta_i[-j]$ (otherwise $T_j$ would have transitioned into $q_A$), $\exists ( 0 \leq i  \leq n ). \alpha_i \not = \beta_i$ (otherwise this would be a prefix of the primary trace),  and $A^j$ accepts $\beta_{p_j}$ (since $T_j $ attempted to transition into a final state and got stuck). Therefore, $\sigma$ does not satisfy the $j$-Deviant-Trace Condition.
\end{proof}

% \begin{proof} We omit this proof in favor of a more comprehensive result presented later on.
% \end{proof}
\subsection{ \emph{W} - NE automata }
We constructed a tree automaton that recognizes the set of strategies that satisfy the Primary-Trace condition for a fixed subset $W \subseteq \Omega$ of agents in an iBG $G$, $T_0=(\Sigma,\Sigma,Q\cup\{q_a\},q_0,\rho_0,F\cup\{q_a\})$.  We also constructed the automaton $T_j$ that checks the $j$-Deviant Trace condition for a specific agent $j$. A simple way to check both the Primary Trace condition and the $j$-Deviant Trace condition for some $W \subseteq \Omega$ would be to take the cross product of $T_0$ with all the $T_j$'s for every $j \not \in W$. We now show that this can be done more efficiently, by taking a modified union of the state sets of $T_0$ and the $T_j$'s instead of their cross product. This is motivated by the observation that each automaton "checks" a disjoint set of paths in a tree $\pi$, and marks all others with a repeating accepting state.

We construct a deterministic top-down B\"uchi infinite-tree automaton $T_W = (\Sigma,\Sigma,Q \cup \bigcup_{j \in \Omega \setminus W} Q^j \cup \{ q_A \} ,q_0,\tau, F \cup \bigcup_{j \in \Omega \setminus W}\{Q^j \setminus F^j\} \cup \{q_A \})$ to accept all strategies that satisfy both the Primary-Trace condition and the $j$-Deviant-Trace Conditions, where
\begin{enumerate}
    \item $\Sigma$ is both the label alphabet of the tree and its set of directions with the $\alpha$ and $\beta$ notations defined as previously.
    \item $q_A$ is a repeating accepting state.
    \item $\tau$ is defined as follows for a given state $q$, label $\alpha$, and  direction $\beta$ 
    \begin{enumerate}
        \item If $q \in Q$
            \begin{enumerate}
                \item 
                If $\alpha = \beta$, then $\tau(q, \alpha, \beta) = \rho_0(q,\alpha, \beta)$
                \item If $\alpha \not = \beta$, but for some $j \in \Omega \setminus W $ we have $\alpha[-j] = \beta[-j]$, 
                then $\tau( q, \alpha, \beta) = \delta^j(q[j], \beta)$, where
                $q[j]$ is $j$-th component of $q$, provided that
                $\delta^j(q[j],\beta)\not\in F^j$.
                \item If for all $j \in \Omega \setminus W$ we have $\alpha[-j] \not = \beta[-j]$, then $\tau(q, \alpha, \beta) = q_A$
            \end{enumerate}
        \item If $q \in Q^j$ for $j\in \Omega \setminus W$, then
            \begin{enumerate}
                %\item If $ q \in F^j , \tau(q, \alpha, \beta) = q$. Otherwise, we check the following.
                %\begin{enumerate}
                \item If $\alpha[-j] = \beta[-j]$, then $\tau(q, \alpha, \beta) = \delta^j(q,\beta)$, provided that $\delta^j(q,\beta)\not\in F^j$,
                \item If $\alpha[-j] \not = \beta[-j]$, then $\tau(q, \alpha, \beta) = q_A$
                %\end{enumerate}
            \end{enumerate}
        \item If $q = q_A$, then $\tau(q, \alpha, \beta) = q_A$
            \end{enumerate} 
\end{enumerate}

Intuitively, the automaton $T_W$ simulates the automaton $T_0$ on the primary trace defined by $\pi$. If the automaton is on the primary trace, it is in a state in $Q$ and it checks all possible $j$-deviations from that state by transitioning accordingly to all states reachable by possible $j$-deviant actions on the corresponding directions. Note that here we only check if $\alpha[-j] = \beta[-j]$ for a single $j$, as it can easy to see that if $\alpha[-j_1] = \beta[-j_1]$ and $\alpha[-j_2] = \beta[-j_2]$ for two different $j_1,j_2$ then $\alpha = \beta$ since $\Sigma_{j_1}$ and $\Sigma_{j_2}$ are disjoint. On a state that does not represent either a continuation of the primary trace or one reachable by a deviation from some agent $j$, we move to the repeating accepting state $q_A$.

If the automaton is in some state $q \in Q^j$, it transitions according to $\delta^j$ on a direction $\beta$ with $\beta[-j] = \alpha[-j]$, including the one where $\alpha = \beta$. On all other directions, it transitions to the new state $q_A$. If the automaton reaches a final state for $A^j$, it gets stuck and cannot accept. This simulates the automaton $T_j$ and verifies the $j$-Deviant-Trace Condition. If the automaton is in the state $q_A$, it means we have marked the subtree starting from the current node as irrelevant to the Nash Equilibrium definition. Therefore, we simply stay in the accepting state $q_A$ on every direction.

\begin{theorem}
Let $G$ be an iBG and $W \subseteq \Omega$ be a set of agents. Let $\pi: \Sigma^*\rightarrow \Sigma$ be a strategy profile. Then $\pi$ is accepted by the tree automaton $T_W$ iff $\pi$ is a $W$-NE strategy.
\end{theorem}
\begin{proof}
% In this proof, all automata $T_0$, $A_W$, and $T_j$ are constructed with respect to $G$ and $W$, and the Primary-Trace Condition and $j$-Deviant Trace Conditions are similarly formulated with respect to $G$ and $W$.

($\rightarrow$)
Suppose $\pi: \Sigma^* \rightarrow \Sigma$ is accepted by $T_W$. We show that $\pi$ must satisfy both the Primary-Trace Condition and the $j$
-Deviant-Trace Condition for all $j\in \Omega \setminus W$.

\begin{enumerate}
    \item The primary trace of $\pi$ is the unique path $p = (\alpha_0 , \beta_0) \ldots $ such that for every $(\alpha_i,\beta_i)$ we have $\alpha_i = \beta_i$. On this path, the automaton $T_W$ stays in states in $Q$ and transitions according to the transition function $\delta$ of $A_W$; thus $T_W$ simulates $A_W$ on the primary trace. Since $T_W$ accepts $\pi$,  we know that $A_W$ accepts on $p$, meaning that exactly the goals $A^i$ for $i \in W$ are satisfied. Therefore $\pi$ satisfies the Primary-Trace Condition.
    \item A $j$-deviant trace of $\pi$ is a path $p_j = (\alpha_0 , \beta_0) \ldots $ such that for every $(\alpha_i,\beta_i) \in p$ we have $\alpha_i[-j] = \beta_i[-j]$ and we have that $p_j$ is different from the primary trace. Therefore, for at least one index $i$, we have that $\alpha_i \not = \beta_i$  in $p_j$. 
    When $T_W$ runs on such a trace, it starts in states in $Q$ and eventually transition to states in $Q^j$ upon reaching the first index where $\alpha_i \not = \beta_i$. When it is in the states in $Q$, $A^j$ cannot reach a final state, otherwise $T_W$ would get stuck and not accept due to the construction of $A_W$, contradicting our assumption that $T_W$ does accept. When it reaches the states in $Q^j$, it also can never get stuck attempting to a transition to final state in $F^j$ due to the construction of the transition function $\tau$, as any such attempted transition would mean $T_W$ would reject. This is true no matter which $j$-deviant trace we choose since $T_W$ accepts on all paths of $\pi$. Therefore $\pi$ satisfies the $j$-Deviant-Trace condition for all $j \in \Omega \setminus W$.
\end{enumerate}

($\leftarrow$)
Note that $T_W$ is deterministic, so there is a unique run $T_W(\pi)$. We have to show that all paths of this run are accepting. There are three types of paths:
\begin{description}
\item{\sf Primary Path:}
If a path $p$ is the primary path, then $T_W$ emulates $A_W$ along $p$. Because of the Primary Trace Condition, we know that $A_W$ eventually enters and stays in the the set $F$ of accepting states. Thus, this path $p$ of $T_W(\pi)$ is accepting.
\item{\sf $j$-Deviant Paths:}
If $p=(\alpha_0,\beta_0),\ldots$ is a $j$-deviant path for some $j\in\Omega\setminus W$, then it can be factored as $p_P \cdot p_j$, with $p_P$ finite, but possibly empty. For every label-direction pair $(\alpha_i,\beta_i)$  in $p_P$ we have that $\alpha_i=\beta_i$ and for every label direction pair $(\alpha_i,\beta_i)$ in $p_j$ we have that $\alpha_i[-j]=\beta_i[-j]$. Note that only one choice of $j$ is appropriate. Let $i$ be the first index in $p$ where $\alpha_i \not = \beta_i$. Having $\alpha_i[-j_1]  = \beta_i[-j_1]$ and  $\alpha_i[-j_2]  = \beta_i[-j_2]$ for two different agents $j_1, j_2$ would imply that $\alpha_i = \beta_i$. $T_W$ first emulates $A_W$ along $p_P$.  Since $\pi$ satisfies the Primary Trace Condition, $T_W$ will never get stuck and reject on $p_P$. Since $p$ is a $j$-Deviant-Trace, there is a smallest $i$ such that $\alpha_i\not=\beta_i$ in $p$. At this point $T_W$ switches from emulating $A_W$ to emulating $A^j$. Because $\pi$ satisfies the $j$-Deviant-Trace-Condition, the goal $A^j$ does not hold along $p$. Thus, $T_W$ does not get stuck along $p_P$ or along $p_j$, and it accepts along $p$.
\item{\sf Other Paths:}
If $p$ is not the primary path nor a $j$-deviant path, then there are two possibilities.
\begin{enumerate}
\item 
The first case is when $p$ can be factored as $p_P \cdot p'$, with $p_P$ finite but possibly empty. For every point $(\alpha_i, \beta_i)$ of $p_P$ we have that $\alpha_i=\beta_i$, and at the first point $(\alpha_k,\beta_k)$ of $p'$ we have that $\alpha_k[-j] \not=\beta_k[-j]$ for all $j\in\Omega \setminus W$. Then $T_W$ will emulate $A_W$ along $p_P$ and transition to $q_A$ upon reading $(\alpha_k,\beta_k)$. By previous arguments, we know that $T_W$ will not get stuck and reject along $p_P$. Once $T_W$ enters $q_A$ it stays in $q_A$, an accepting state. Therefore $T_W$ accepts the path $p =p_P \cdot p' $ 
\item
The second case is when $p$ can be factored as $p_P \cdot p_j \cdot p'$, with $p_P$ finite but possibly empty  and $p_j$ finite and nonempty. For every label-direction pair $(\alpha_i,\beta_i)$ in $p_P$ we have that $\alpha_i=\beta_i$. For some $j\in\Omega\setminus W$ we have that $\alpha_i[-j]=\beta_i[-j]$ for every label-direction pair $(\alpha_i,\beta_i)$ in $p_j$, again noting that only one choice of $j$ is appropriate. Finally, at the first point $(\alpha_k,\beta_k)$ of $p'$ we have that $\alpha_k[-j] \not=\beta_k[-j]$.  By previous arguments, we know that $T_W$ will not get stuck and reject along $p_P$ or $p_j$. And since $T_W$ transitions to $q_A$ at the beginning of $p'$, we know that it cannot get stuck and reject along $p'$. Therefore $T_W$ will accept on $p = p_P \cdot p_j \cdot p'$.
\end{enumerate}
\end{description}
\end{proof}

\begin{corollary}
Let $G$ be an iBG and $W \subseteq \Omega$ be a set of agents. Then, a $W$-NE strategy exists in $G$ iff the automaton $T_W$ constructed with respect to $G$ is nonempty.
\end{corollary}

\section{Algorithmic Framework}
In the previous section, we constructed an automaton $T_W$ that recognizes the set of Nash equilibrium strategy profiles with winning set $W$ in an iBG $G$, which we denoted as $W$-NE strategies. The problem of determining whether a $W$-NE strategy exists is equivalent to testing $T_W$ for nonemptiness. The standard algorithm for testing nonemptiness of B\"uchi tree automata involves B\"uchi games \cite{GTW02}. In this section, we prove that testing $T_W$ for nonemptiness is equivalent to solving safety games and then testing a B\"uchi word automata for nonemptiness. This gives us a simpler path towards constructing an optimal algorithm to decide our central question.

\subsection{Safety Game for Deviating Agents}
Note that the B\"uchi condition on the j-Deviant traces simply consists of avoiding the set of final states in $A^j$, making it simpler than a general B\"uchi acceptance condition. In order to characterize this condition precisely, we now construct a 2-player safety game that partitions the states of $Q^j$ in $T_W$, for $j\in\Omega\setminus W$, into two sets -- states in which $T_W$ started in state $q \in Q^j$ is empty and states in which $T_j$ started in state $q \in Q^j$ is nonempty. 
%Recall the DFA goal for an agent $j$ is given by $A^j=(2^{Prop},Q^j,q^j_0,\delta^j,F^j)$.
We construct the safety  game $G_j = (Q^j, Q^j \times \Sigma, E_j)$. The safety set can intuitively be thought of as all the vertices not in $F^j$, but for our purposes it is more convenient to not define outgoing transitions from these states - thus making them losing for Player 0 by violating the infinite play condition. Player 0 owns $Q^j$ and Player 1 owns $Q^j\times \Sigma$. Here we retain our $\alpha$ and $\beta$ notation in so far as they are both elements of $\Sigma$. The edge relation $E_j$ is defined as follows:
\begin{enumerate}
\item 
$(q,\langle q,\alpha\rangle)) \in E_j$ for $q \in Q^j\setminus F^j$ and $\alpha \in \Sigma$.
\item 
$(\langle q,\alpha\rangle, q') \in E_j$ for $q\in Q^j$ and $q'\in Q^j$, where $q'=\delta^j(q,\beta)$ for some $\beta\in \Sigma$ such that $\alpha[-j]=\beta[-j]$.
\end{enumerate}
As defined above, if $q\in F^j$, then $q$ has no successor node, and Player 0 is stuck and loses the game. Since $G_j$ is a safety game, Player 0's goal is to avoid states in $F^j$ and not get stuck. Let $Win_0(G_j)$ be the set of winning states for Player 0 in the safety game $G_j$.

\begin{theorem}\label{Win}
 A state $q\in Q^j\setminus F^j$ belongs to $Win_0(G_j)$ iff $T_W$ is nonempty when started in state $q$.
\end{theorem}

\begin{proof}
($\rightarrow)$
Suppose $q \in Q^j\setminus F^j$ and $q \in Win_0(G_j)$. We construct a tree $\pi_q : \Sigma^* \rightarrow \Sigma$ that is accepted by $T_W$ starting in state $q$. To show that $\pi_q$ is accepted, we also construct an accepting run $r_q : \Sigma^* \rightarrow (Q^j\setminus F^j) \cup\{q_A\}$. By construction, we have $r_q(x)\in Win_0(G_j)$ for all $x\in \Sigma^*$. We proceed by induction on the length of the run.

For the basis of the induction, we start by defining $\pi_q(\varepsilon)$ and $r_q(\varepsilon)$. First, we let $r_q(\varepsilon)=q$. By the assumption that $q \not \in F^j$, the run cannot get stuck and reject here.

For the step case, suppose now that we have constructed  $r_q(y)=p\in Win_0(G_j)$ for some $y\in \Sigma^*$. Now, since $p\in Win_0(G_j)$ and  cannot get stuck, there must be a node $\langle q, \alpha_y \rangle$ contained in both $Q^j \times \Sigma$  and $Win_0(G_j)$, so we let $\pi_q(y)=\alpha_y$.
 Recall that the directions of $\pi$ are $\Sigma$. Divide the possible directions $\beta\in \Sigma$ into two types: either $\alpha_y[-j] = \beta[-j]$ or $\alpha_y[-j] \not = \beta[-j]$.  If $\alpha_y[-j] = \beta[-j]$, then this corresponds to a legal move by player 1 in $G_j$. Since $\langle q, \alpha_y \rangle\in Win_0(G_j)$, moves by player 1 must stay in $Win_0(G_j)$. It follows that $q'=\delta^j(q,\beta)\in Win_0(G_j)$, so $q'\not\in F^j$. We let $r_q(y \cdot \beta)=q'$. If, on the other hand, $\alpha_y[-j] \not= \beta[-j]$, we let $r_q(y\cdot \beta)=q_A$. Once we have reached a node  $z\in \Sigma^*$ with $r_q(z)=q_A$, we define $r_q(z')=q_A$ for all descendants $z'$ of $z$ and we can define $\pi_q(z')$ arbitrarily. Since we can never get stuck, we never reach a state in $F^j$, so the run $r_q$ is accepting.

$(\leftarrow)$
Suppose now that $T_W$ started in state $q$ accepts a tree $\pi_q : \Sigma^* \rightarrow \Sigma$. Since the automaton $T_W$ is deterministic, it accepts with a unique run of $T_W$ on $\pi_q$ as $r_q: \Sigma^* \rightarrow (Q^j\setminus F^j) \cup\{q_A\}$. We claim that $\pi_q$ is a winning strategy for player 0 in $G_j$ from the state $q$.  

Consider a play $\pi=p_0,\alpha_0,\beta_0,p_1,\alpha_1,\beta_1,\ldots$, where $p_i\in Q^j$, $p_0=q$, and $\alpha_i,\beta_i \in \Sigma$. In round $i\geq 0$, player 0 moves from $p_i$ to $\langle p_i, \alpha_i \rangle$, for $\alpha_i=\pi_q (\langle \beta_0,\ldots,\beta_{i-1} \rangle)$, and then player 1 moves from $\langle p_i, \alpha_i\rangle$ to $p_{i+1}=\delta^j(p_i,\beta_i)$, for some $\beta_i$ such that $\alpha_i[-j]=\beta_i[-j]$. Let $x_i=\langle \beta_0,\ldots,\beta_{i-1} \rangle$, so we have that $\alpha_i=\pi_q(x_i)$. By induction on the length of $x_i$ it follows that $p_i=r_q(x_i)$. Since $r_q$ is an accepting run of $T_W$ on $\pi_q$, it follows that $p_i=r_q(x_i)\not\in F^j$. Thus, the play $\pi$ is a winning play for player 0. It follows that $\pi_q$ is a winning strategy for player 0 in $G_j$ from the state $q$.
\end{proof}

\subsection{A B\"uchi Automaton for $T_W$ Nonemptiness}

Recall that the tree automaton $T_W$, which recognizes $W$-NE strategies, emulates the B\"uchi automaton $A_W=(Q,q_0,\Sigma,\delta,F)$ along the primary trace and the goal automaton $A^j$ along $j$-deviant traces. We have constructed the above games $G_j$ to capture nonemptiness of $T_W$ from states in $Q^j$, in terms of the winning sets $Win_0(G_j)$. We now modify $A_W$ to take these safety games into account. Let $A'_W=(Q',q_0,\Sigma,\delta',F \cap Q')$ be obtained from $A_W$ by restricting states to $Q'\subseteq Q$, where $Q' = \bigtimes_{i \in W} Q^i \times \bigtimes_{j\in \Omega\setminus W} \{ Win_0(G_j) \cap Q^j \} \times 2^{\Omega}$. In other words, the $j^{th}$-component $q_{i_j}$ of a state $\overline{q}=\langle q_{i_1},\ldots,q_{i_n}\rangle \in Q'$ must be in $Win_0(G_j)$ for all $j \in \Omega \setminus W$, otherwise the automaton $A'_W$ gets stuck. Finally, we define the transition function $\delta'$ as follows:
$\delta'(q,\sigma)=\delta(q,\sigma)$ if, for all $j \not \in W$, we have that $(q[j], \sigma) \in Win_0(G_j)$; otherwise, $\delta'(q,\sigma)$ is undefined. Intuitively, the letter $\sigma$ must be a winning move for Player 0 in the safety game $G_j$.
%\begin{itemize}
%    \item $\delta(q,\sigma)$ if for all $j \not \in W$ $\delta^j(q[j],\sigma) %\in Win_0(G_j) $ and $(q[j], \sigma) \in Win_0(G_j)$
%    \item undefined otherwise
%\end{itemize} 
%    \item undefined otherwise
%\end{itemize}
%\begin{itemize}
%    \item $\delta(q,\sigma)$ if for all $j \not \in W$ $\delta^j(q[j],\sigma) %\in Win_0(G_j) $ and $(q[j], \sigma) \in Win_0(G_j)$
%    \item undefined otherwise
%\end{itemize}
%Intuitively, we check to make sure that we never leave the winning set for Player 0 when considering transitioning over a letter $\sigma$ to ensure that these agents can never succesfully deviate.

\begin{theorem}
The B\"uchi word automaton $A'_W$ is nonempty iff the tree automaton $T_W$ is nonempty.
\end{theorem}

\begin{proof}
($\rightarrow$)
Assume $A'_W$ is nonempty. Then, it accepts an infinite word $w=w_0 w_1 \ldots \in \Sigma^{\omega}$ with a run $r=q_0, q_1, \ldots\in {Q'}^{\omega}$. We use $w$ and $r$ to create a tree $\pi :\Sigma^* \rightarrow \Sigma $ with an accepting run $r_\pi: \Sigma^* \rightarrow Q\cup \{q_A\}$ with respect to $T_W$.

Let $x_0=\varepsilon$. We start by setting $\pi(x_0)=w_0$ and $r_{\pi}(x_0)=q_0$. Suppose now that we have just defined $\pi(x_i)=\alpha$ and $r_\pi(x_i)=q$, and, by construction, $x_i$ is on the primary trace. Consider now the node $x_i \cdot \beta$, $\beta \in \Sigma$. There are three cases to consider:
\begin{enumerate}
\item
If $\pi(x_i)=\beta$, then we set $x_{i+1}= x_i \cdot \beta$, $\pi(x_{i+1})=w_{i+1}$ and $r_\pi(x_{i+1})=q_{i+1}$. Note that $x_{i+1}$ is, by construction, the successor of $x_i$ on the primary trace. Thus, the projection of $r_\pi$ on the primary trace of $\pi$ is precisely $r$, so $r_\pi$ is accepting along the primary path.
\item
If  $\pi(x_i)[-j]=\beta[-j]$ and $\pi(x_i) \not =\beta$ for some $j\in\Omega\setminus W$, then we set $r_\pi(x_i \cdot \beta)=q'_j=\delta^j(q_j,\beta)$, where $q_j$ is the $j$-th component of $q$. Since $(q_j, \pi(x_i)) \in Win_0(G_j))$ (otherwise, the transition would not be defined), we have that ${q'_j} \in Win_0(G_j)$.
By Theorem~\ref{Win}, $T_W$ is nonempty when started in state ${q'_j} $. That is, there is a tree $\pi_{q'_j} $ and an accepting run $r_{q'_j} $ of $T_W$ on $\pi_{q'_j} $, starting from ${q'_j}$. So we take the subtree of $\pi$ rooted at the node $x_i \cdot \beta$ to be $\pi_{q'_j} $, and the run of $T_W$ from $x_i \cdot \beta$ is $r_{q'_j}$. So all paths of $r_\pi$ that go through $x_i \cdot \beta$ are accepting.
\item
Finally, if $\pi(x_i)[-j]\not =\beta[-j]$ for all $j\in\Omega\setminus W$, then $x_i \cdot \beta$ is neither on the primary trace nor on a $j$-deviant trace for some $j\in\Omega\setminus W$. So we set $r_\pi(x_i \cdot \beta)=q_A$ as well as $r_\pi(y)=q_A$ for all descendants $y$ of $x_i \cdot \beta$. The labels of $x_i \cdot \beta$ and it descendants can be set arbitrarily. So all paths of $r_\pi$ that go through $x_i \cdot \beta$ are accepting.

\end{enumerate}

($\leftarrow$)
Assume $T_W$ is nonempty. Then, we know that it accepts at least one tree $\pi : \Sigma^* \rightarrow \Sigma$. In particular, since $T_W$ accepts on all branches of $\pi$ it accepts on the primary trace, denoted as $\pi_p$.

Since $T_W$ accepts on $\pi_p$, we can consider the run of $T_W$ on $\pi$ which we denote $r : \Sigma^* \rightarrow Q$. Let the image of $r(\pi_p)$ be $Q^* \subseteq Q$. We claim that $Q^* \subseteq Q'$.

Assume otherwise, that for some finite prefix of the primary trace of $\pi$ denoted $p$ we have that $r(p) \not \in Q'$. Since $r(p)$ clearly is inside $Q$, it must be the case that for some $j \in \Omega \setminus W$ $r(p)[j] \not \in Win_0(G_j)$. Since $r(p)[j]$ is not in $Win_0(G_j)$, it must be in $Win_1(G_j)$. This means that, upon observing $p$, a direction $\beta$ exists that transitions $T_W$ into a state $q' \in Win_1(G_j)$. From here player 1 has a winning strategy in $G_j$. Following one of the paths created by player 1 playing directions according to this winning strategy and player 0 playing anything in response, we get that player 1 will eventually win the game, forcing $T_W$ to attempt a transition into $F^j$ and getting stuck. Therefore $T_W$ does not actually accept $\pi$, a contradiction.

Since the image of $r(\pi_p)$ is contained within  $Q'$, we claim that $A'_W$ accepts the word formed by the labels along $\pi_p$, which we denote by $\alpha(\pi_p)$. Since $T_W$ accepts along $\pi_p$ and the run $r(\pi_p)$ never leaves $Q'$, we have that there are infinitely many members of the set $F \cap Q'$ in the run $r(\pi_p)$, satisfying the B\"uchi condition of $A'_W$. Furthermore, since a winning state is never reached for an agent $j \not \in W$ in $G_j$, we have that we never are a node labeled $\sigma$ in a state containing $q_j$ such that $(q_j,\sigma) \in Win_1(G_j)$ - otherwise a state in $Win_1(G_j)$ could be reached. This implies that all labels correspond to possible transitions in $A'_W$.  And since any states in which some $Q^j$ for $j \in \Omega \setminus W$ reaches a final state are excluded from $Q'$, $A'_W$ will never get stuck reading $\alpha(\pi_p)$. Therefore, $A'_W$ accepts  $\alpha(\pi_p)$ and is therefore nonempty.

\end{proof}

\begin{corollary}
Let $G$ be an iBG and $W \subseteq \Omega$ be a set of agents. Then, a $W$-NE strategy exists in $G$ iff the automaton $A'_W$ constructed with respect to $G$ is nonempty.
\end{corollary}

\section{Complexity and Algorithms}

\subsection{Complexity}
The algorithm outlined by our previous constructions consists of two main part. First, we construct and solve a safety game for each agent. Second, for $W \subseteq \Omega$, we check the automaton $A'_W$ for nonemptiness. The input to this algorithm consists of $k$ goal DFAs with alphabet $\Sigma$ and a set of $k$ alphabets $\Sigma_i$ corresponding to the actions available to each agent. Therefore, the size of the input is the sum of the sizes of these $k$ goal DFAs.

In the first step, we construct a safety game for each of the agents. The size of the state space of the safety game for agent $j$ is $|Q^j|(|\Sigma|+1)$. The size of the edge set for the safety game can be bounded by $(|Q^j|*|\Sigma|) + (|Q^j|^2*|\Sigma|)$, where $|Q^j|*|\Sigma|$ represents the $|\Sigma|$ outgoing transitions from each state in $Q^j$ owned by player $0$ and $|Q^j|^2*|\Sigma|$ is an upper bound assuming that each of the states in $Q^j \times \Sigma$ owned by player $1$ can transition to each of the states in $Q^j$ owned by player $0$. Since safety games can be solved in linear time with respect to the number of the edges \cite{bernet2002}, each safety game is solved in polynomial time. We solve one such safety game for each agent which represents a linear blow up. Therefore, solving the safety games for all agents can be done in polynomial time.

For a given $W \subseteq \Omega$, querying the automaton $A'_W$ for nonemptiness can be done in PSPACE, as the state space of $A'_W$ consists of tuples from the product of input DFAs. We can then test $A'_W$ on the fly by guessing the prefix of the lasso and then guessing the cycle, which can be done in polynomial space \cite{VW94}.

\begin{theorem}
The problem of deciding whether there exists a $W$-NE strategy profile
for an iBG $G$ and a set $W \subseteq \Omega$ of agents is in PSPACE.
\end{theorem}

\subsection{PSPACE Lower Bound}
In this section we show that the problem of determining whether a $W$-NE exists in an iBG is PSPACE-hard by providing a reduction from the PSPACE-complete problem of DFA Intersection Emptiness (DFAIE). The DFAIE problem is as follows: Given $k$ DFAs $A^0 \ldots A^{k-1}$ with a common alphabet $\Sigma$, decide whether $\bigcap_{0 \leq i \leq k-1}A^i \not = \emptyset$ \cite{Kozen77}.

Given a DFA $A^i=\langle Q^i, q^i_0 , \Sigma, \delta^i,  F^i \rangle$, we define the goal DFA $\hat{A}^i=\langle \hat{Q}^i, q^i_0 , \hat{\Sigma}, \hat{\delta}^i,  \hat{F}^i \rangle $ as follows: 
\begin{enumerate}
    \item $\hat{\Sigma}=\Sigma \cup \{ K \}$, where $K$ is a  new symbol, i.e. $K \not\in \Sigma$
    \item $\hat{Q}^i=Q^i\cup\{ \accept, \reject \}$,
    \item 
    \begin{equation*}
    \begin{cases}
    \hat{\delta}^i(q,a)=q \mbox{ for } q\in \{\accept,\reject\} \mbox{ and } a\in \hat{\Sigma}\\ 
    \hat{\delta}^i(q,a)=\delta^i(q,a) \mbox{ for } q\in Q^i \mbox{ and } a\in\Sigma\\
    \hat{\delta}^i(q,K)= \accept \mbox{ for } q\in F^i\\
    \hat{\delta}^i(q,K)= \reject \mbox{ for } q\in Q^i \setminus F^i
    \end{cases}
    \end{equation*}
        \item $\hat{F}^i=\{ \accept \}$
   \end{enumerate}
Intuitively, $\accept$ and $\reject$ are two new accepting and rejecting states that have no outgoing transitions. The new symbol $K$ takes accepting states to $\accept$ and rejecting states to $\reject$. The purpose of $K$ is to synchronize acceptance by all goal automata. We call the process of modifying $A^i$ into $\hat{A^i}$ transformation.

The transformation from $A^i$ to $\hat{A}^i$ can be done in linear time with respect to the size of $A^i$, as the process only involves adding two new states. Furthermore, if $A^i$ is a DFA then $\hat{A}^i$ is also a DFA. 

Given an instance of DFAIE, i.e., $k$ DFAs $A^0 \ldots A^{k-1}$, we create an iBG $G$, defined in the following manner. 
\begin{enumerate}
    \item $\Omega = \{ 0, 1 \ldots k-1 \}$
    \item The goal for agent $i$ is $\hat{A}^i$
    \item $\Sigma_0 = \Sigma \cup \{K\}$
    \item $\Sigma_i = \{*\}$ for $i \not= 0$. Here $*$ represents a fresh symbol, i.e., $* \not \in \Sigma$ and $* \not = K$.
\end{enumerate}

Clearly, the blow-up of the construction is linear. Since each agent except $0$ is given control over a set consisting solely of $*$, the common alphabet of the $\hat{A}^i$ is technically $\hat{\Sigma} \times \{ * \}^{k-1}$. This alphabet is isomorphic to $\hat{\Sigma}$, so by a slight abuse of notation we keep considering the alphabet of the $\hat{A}^i$ to be $\hat{\Sigma}$.

Before stating and proving the correctness of the reduction, we make two observations. We are  interested here in Nash equilibria in which \emph{every} agent is included in $W$. This implies the following:
\begin{enumerate}
    \item The existence of an $\Omega$-NE is defined solely by the Primary-Trace Condition. Since there are no agents in $\Omega \setminus W$, there is no concept of a $j$-Deviant-Trace. If we are given an infinite word that satisfies the Primary-Trace Condition, we can extend it to a full $\Omega$-NE strategy tree by labeling the nodes that do not occur on the primary trace arbitrarily.
    \item Since there are no $j$-Deviant-Traces in this specific instance of the $\Omega$-NE Nonemptiness problem, we can relax our assumption that $|\Sigma_j| \geq 2$ for all $j\in \Omega$, since there is no meaningful concept of deviation in an $\Omega$-NE. Recall that this assumption was made only for simplicity of presentation regarding $j$-Deviant Traces.

\end{enumerate}

\begin{theorem}
Let $A^0\ldots A^{k-1}$ be $k$ DFAs with alphabet $\Sigma$. Then, $\bigcap_{0 \leq i \leq k-1} L(A^i) \not= \emptyset$ iff there exists an $\Omega$-NE in the iBG $G$ constructed from  $A^0\ldots A^{k-1}$. 
\end{theorem}

\begin{proof}
In this proof, we introduce the notation $S$ to denote an infinite suffix, which is an arbitrarily chosen element of  $\{ \Sigma \cup K \}^{\omega}$. 

$(\rightarrow)$
Assume that $\bigcap_{0 \leq i \leq k-1} L(A^i) \not= \emptyset$. Then, there is a word $w \in \Sigma^*$ that is accepted by each of $A^0 \ldots A^{k-1}$. We now show that $w \cdot K \cdot S$ satisfies all goals $\hat{A}^0\ldots \hat{A}^{k-1}$. Since each of $A^0\ldots A^{k-1}$ accepts $w$, each of $\hat{A}^0 \ldots \hat{A}^{k-1}$ reaches a final state of $A^0 \ldots A^{k-1}$, respectively, after reading $w$. Then, after reading $K$, $\hat{A}^0\ldots \hat{A}^{k-1}$ all simultaneously transition to $\accept$. Therefore all goals $\hat{A}^i$ are satisfied on $w \cdot K \cdot S$ and  $w \cdot K \cdot S$ satisfies the Primary-Trace Condition. Since we are considering an $\Omega$-NE, there is no need to check deviant traces and $w \cdot K \cdot S$  can be arbitrarily extended to a full $\Omega$-NE strategy profile tree. 

$(\leftarrow)$
Assume that the iBG $G$ with goals $\hat{A}^0\ldots \hat{A}^{k-1}$ admits an $\Omega$-NE. We claim that its primary trace must be of the form $w \cdot K \cdot S$, where $w \in \Sigma^*$ does not contain $K$. This is equivalent to saying that a satisfying primary trace must have at least one $K$. This is easy to see, as the character $K$ is the only way to transition into an accepting state for each $\hat{A}^i$, therefore it must occur at least once if all $\hat{A}^i$ are satisfied on this trace.

We now claim that each of $A^0\ldots A^{k-1}$ accept $w$. Assume this is not the case, and some $A^i$ does not accept $w$. Then, while reading $w$, $\hat{A}^i$ never reaches $\accept$, as $w$ does not contain $K$. Furthermore, upon seeing the first $K$, $\hat{A}^i$ transitions to $\reject$, since $A^i$ is not in a final state in $F^i$ after reading $w$. Thus, $\hat{A}^i$ can never reach $\accept$, contradicting the assumption that $w \cdot K \cdot S$ was an $\Omega$-NE. Therefore all $A^i$ must accept $w$, and $\bigcap_{0 \leq i \leq k-1} L(A^i) \not = \emptyset$.

\end{proof}
This establishes a polynomial time reduction from DFAIE to $W$-NE Nonemptiness; therefore $W$-NE Nonemptiness is PSPACE-hard. In fact this reduction has shown that checking the Primary-Trace Condition is itself PSPACE-hard. Combining this with our PSPACE decision algorithm yields PSPACE-completeness.

\begin{theorem}
The problem of deciding whether there exists a $W$-NE strategy profile
for an iBG $G$ and a set $W \subseteq \Omega$ of agents is PSPACE-complete.
\end{theorem}

\section{Concluding Remarks}
The main contribution of this work is Theorem 5.3, which characterizes the complexity of deciding whether a $W$-NE strategy profile exists for an iBG $G$ and $ W \subseteq \Omega$ is PSPACE-complete.

\paragraph{Separation of Strategic and Temporal Reasoning}: The main objectives of this work is to analyze equilibria in finite-horizon multiagent concurrent games, focusing on the strategic-reasoning aspect of the problem, separately from temporal reasoning. In order to accomplish this, we used DFA goals instead of goals expressed in some finite-horizon temporal logic. For these finite-horizon temporal logics, previous analysis \cite{GPW17} consisted of two steps. First, the logical goals are translated into a DFA, which involves a doubly exponential blow up \cite{GV13,KV01d}. The second step was to perform the strategic reasoning, i.e., finding the Nash equilibria with the DFA from the first step as input. In terms of computational complexity, the first step completely dominated the second step, in which the strategic reasoning was conducted with respect to the DFAs. Here we eliminated the doubly exponential-blow up from consideration by starting with DFA goals and provided a PSPACE-completeness result for the second step.

% \paragraph{Finitary Fixpoint-Based Computation}
% In addition to the PSPACE completeness result, we also presented another algorithm for solving the problem that was solely comprised of fixpoint subroutines. We created this algorithm with implementation in mind, especially symbolic implementations. Recently, work has been done to represent formulas from finite horizon temporal logics such as LTL\textsubscript{f} as fully symbolic automata \cite{ZTLPV17}. With these advances in mind, a fixpoint-based algorithm is a natural choice to act upon the fully symbolic outputs of these algorithms. We also believe that presenting the algorithm in terms of safety and reachability queries makes it easier to understand,  more scalable, and ultimately more efficient.

\paragraph{Future Work}: Our immediate next goals are to analyze problems such as verification (deciding whether a given strategy profile is a $W$-NE) and strategy extraction (i.e., construction a finite-state controller that implements the $W$-NEs found) within the context of our DFA based iBGs. Furthermore, we are interested in implementation, i.e. a tool based on the theory developed in this paper. Further points of interest can be motivated from a game-theory lens, such as introducing imperfect information. Earlier work has already introduced imperfect information to problems in synthesis and verification - see \cite{BMMRV18, GV16,TabaVar20}. Finally, the work can be extended to both the general CGS formalism (as opposed to iBGs) and to querying other properties/equilibrium concepts outside of the Nash equilibria. \emph{Strategy Logic} \cite{MMPV14} has been introduced as a way to query general game theoretic properties on concurrent game structures, and a version of strategy logic with finite goals would be a promising place to start for these extensions.

\bibliographystyle{abbrv} 
\bibliography{references}

\pagebreak
\begin{figure*}[t]
\begin{tabular}{ |p{1.5cm}||p{15cm}|  }
 \hline
 \multicolumn{2}{|c|}{Notation Glossary} \\
 \hline
Symbol & Meaning\\
\hline
$G$ & An iBG as a whole  \\
$\Omega$ & The set of agents in an iBG.   \\
$k$ & The cardinality of $\Omega$. $\Omega = \{ 0 ,1 \ldots k-1 \}$ \\ 
$i,j$ & Agents in an iBG. $j$ usually refers to a deviating agent.\\
$A^i$ & The goal automata for agent $i$. Usually $A^j$ if $j$ is not in $W$. $A^i = \langle Q^i, q^i_0, \Sigma, \delta^i ,F^i \rangle$\\
$\pi_i$ & A strategy for agent $i$ \\
$\Pi_i$ & The set of strategies for agent $i$\\
$\pi$ & A strategy profile consisting of one strategy for each agent. $\pi = \langle \pi_0 \ldots \pi_{k-1} \rangle$. In the context of safety games, a play. \\ 
$\Sigma_i$ & The set of actions for agent $i$\\
$\Sigma$& The cross product of all $\sigma_i$. \\
$w$ & An element of $\Sigma^*$\\
$\alpha$ & An element of $\Sigma$. In the context of $\Sigma^* \rightarrow \Sigma$ trees, it refers to a label.\\
$\beta$ & An element of $\Sigma$. In the context of $\Sigma^* \rightarrow \Sigma$ trees, it refers to a direction.\\
$q_A,q_a$ & A catch all accepting state in tree automata that always transitions back to itself.\\ 
$A_W$    & A deterministic B\"uchi word automaton that accepts traces in which all goals from $W$ are satisfied and no others. $A_W = \langle Q, q_0, \Sigma, \delta, F \rangle$  \\
$T_0$ & A deterministic top down B\"uchi tree automaton that accepts a tree if its primary trace is accepted by $A_W$. $T_0 =(\Sigma,\Sigma,Q\cup\{q_a\},q_0,\rho_0,F\cup\{q_a\}) $  \\
$T_j$ & A deterministic top down B\"uchi  tree automaton that accepts a tree if it satisfies the j-Deviant Trace Condition. $T_j=(\Sigma,\Sigma, (Q^j \times \{0,1\}) \cup \{q_A \} ,\langle q^j_0,0\rangle,\rho_j,(Q^j \times \{0\}) \cup ((Q^j \setminus F^j)\times \{1\}) \cup \{q_A \})$\\
$T_W$ &  A deterministic top down B\"uchi  tree automaton that accepts a tree if it represents a $W$-NE strategy profile. $T_W = (\Sigma,\Sigma,Q \cup \bigcup_{j \in \Omega \setminus W} Q^j \cup \{ q_A \} ,q_0,\tau, F \cup \bigcup_{j \in \Omega \setminus W}\{Q^j \setminus F^j\} \cup \{q_A \})$ \\
$G_j$ & A safety game constructed to characterize the states of $Q^j$ in $A^j$ into those that $T_W$ started in said state is empty or not. $G_j = (Q^j, Q^j \times \Sigma, E_j)$\\
$Win_0(G_j)$ & The winning set of player $0$ in $G_j$.\\
$A'_W$ & A deterministic B\"uchi word automata used to test $T_W$ for nonemptiness. $A'_W=(Q',q_0,\Sigma,\delta',F \cap Q')$\\
$K$ & A fresh character that is not contained in $\Sigma$.\\
$*$ & A second fresh character that is neither contained in $\Sigma$ nor equal to $K$.\\
$\hat{A}^i$ & A transformed DFA that serves as a goal DFA in an iBG. $\hat{A}^i=\langle \hat{Q}^i, q^i_0 , \hat{\Sigma}, \hat{\delta}^i,  \hat{F}^i \rangle $ See Section 5.2.\\
$\hat{\Sigma}$ & $\Sigma \cup \{ K \}$\\
$S$ & An arbitrary element of $\{ \Sigma \cup K \} ^{\omega}$\\
 \hline
\end{tabular}
\end{figure*}
\end{document}